\documentclass[5p, sort&compress]{elsarticle}

\usepackage{lineno,hyperref}

\usepackage{amsmath}

\journal{Nuclear Instruments and Methods in Physics Research Section A}

\usepackage{url}
\biboptions{sort,compress}

\usepackage[squaren]{SIunits}
\newcommand{\um}{\micro\meter}
\newcommand{\mm}{\milli\meter}

\begin{document}

\begin{frontmatter}

\title{High resolution MCP-PMT Readout Using Transmission Lines}


\author[IRFU]{M.~Follin} 
\author[IRFU]{R.~Chyzh}
\author[IRFU]{C.-H.~Sung}
\author[IJCLab]{D.~Breton} 
\author[IJCLab]{J.~Maalmi}
\author[IRFU]{T.~Chaminade}
\author[IRFU]{E.~Delagnes}
\author[Munster-EIMI]{K.~Sch\"afers}
\author[Munster-NPI]{C.~Weinheimer}
\author[IRFU,BioMaps]{D.~Yvon}
\author[IRFU,BioMaps]{V.~Sharyy\corref{correspondingauthor}}

\address[IRFU]{IRFU, CEA,  Universit\'e Paris-Saclay,  Gif-sur-Yvette, France}
\address[BioMaps]{BioMAPs, Service Hospitalier Fr\'ed\'eric Joliot, CEA, CNRS, Inserm, Universit\'e Paris-Saclay, Orsay, France}
\address[IJCLab]{IJCLab, IN2P3, CNRS, Universit\'e Paris-Saclay, Orsay, France}
\address[Munster-NPI]{Nuclear Physics Institute, University of M\"unster, M\"unster, Germany}
\address[Munster-EIMI]{European Institute for Molecular Imaging, University of M\"unster, M\"unster, Germany}
 
\cortext[correspondingauthor]{Corresponding author}
\ead{viatcheslav.sharyy@cea.fr}


\begin{abstract}

  In this article we study the potential of the MCP-PMT read-out to detect  single photo-electron
  using transmission lines. Such a solution limits
  the number of read-out channels, has a uniform time resolution across the PMT surface and provides quasi-continuous
  measurement of the spatial coordinates. The proposed solution is designed to be used in the BOLD-PET project aiming to
  develop an innovative detection module for the positron emission tomography
  using the liquid detection media, the tri-methyl bismuth.
  In this study we use the commercial MCP-PMT Planacon from Photonis, with 32x32 anode structure. The PCB gathers
  signals from anode pads in 32 transmission lines which are read-out from both ends.
  Amplifier boards and SAMPIC modules, developed
  in our labs, allow us to perform the cost-effective, multi-channel digitization of signals with excellent precision.
  For a single photo-electron, we measured a time resolution of 70~ps (FWHM) simultaneously with a spatial accuracy
  of 1.6~mm and 0.9~mm (FWHM) along and across transmission lines correspondingly.

\end{abstract}


\end{frontmatter}

\section{Introduction}
Micro-channel-plate photo multiplier tubes (MCP-PMTs) are position sensitive PMTs which provide the best time resolution up to now.
MCP-PMT has quite large surface, typically 60~mm x 60~mm \citep{Photonis_PMT,Hamamatsu_PMT, Photek_PMT}
and even up to 200 mm x 200 mm~\citep{Minot2019Aug}.
The single photo-electron time transit spread (TTS) is of about
40~ps (SD\footnote{Standard Deviation})~\citep{Vavra2020Apr,Canot2019Dec}
In addition, MCP-PMT has a low dark count rate of about 100  Hz/cm$^2$ (depending on photocathode technology),
which makes it an excellent choice for detecting the Cherenkov radiation, for example, in particle physics
experiments~\citep{Vavra2020Apr,PandaCollaboration2019Dec,Brook2018May,Grigoryev2016,Trzaska2016,Hirose2015Jul}.
In order to allow an accurate coordinate resolution, the PMT anode is often structured in 32 x 32 or 64 x 64 pads,
providing a resolution of the order of 1~mm and better.
The large number of anodes requires a large number of electronics channels if read-out individually.
To reduce the number of channels a read-out scheme based on delay lines is used.
Most of these configurations are optimized for obtaining a high coordinate resolution, see e.g.~\citep{Siegmund1993Nov,Timothy2013},
but in the project BOLD-PET we need to achieve simultaneously a high time resolution.

\begin{figure}
    \centering\includegraphics[width=.36\textwidth]{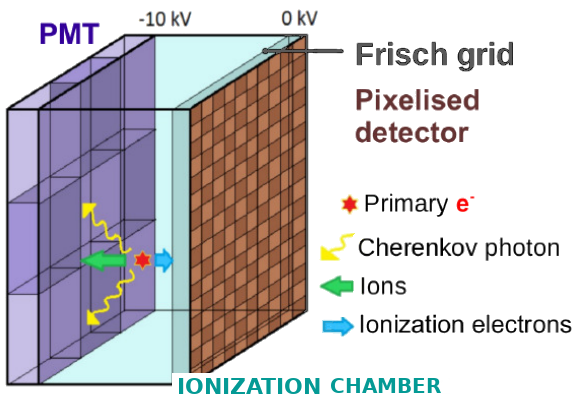}
    \caption{\label{fig:BOLDPET} Principle scheme of the BOLD-PET detector.
      The primary electron created by 511-keV $\gamma$ produces Cherenkov light and ionizes the medium.}
\end{figure}

The BOLD-PET projects aims to develop an innovative detection module for precise positron emission tomography (PET) imaging.
The detection module uses a heavy liquid media,
tri-methyl bismuth (TMBi), which has a short, 25 mm, attenuation length and high, 47\%,
photoelectric fraction for converting the 511~keV photon
to a relativistic electron.
This  electron produces Cherenkov radiation  and ionizes the medium.
The detector operates as a time-projection chamber and detects both light and charge
signals, see Fig.~\ref{fig:BOLDPET} and \citep{Yvon2014, Ramos2015}.
Ionization signal is read-out in  a classical Frisch grid configuration using densely pixelated anode matrix
and is used to determine precisely the 2D position of the gamma interaction. 
The optical photons provide a precise timing to determine the 3D coordinate using the ionization drift time and 
to improve the quality of the image reconstruction with the time-of-flight technique~\citep{Vandenberghe_2016,Lecoq2020Oct,Schaart2021Mar}.
In this project, the optical photons are detected with a MCP-PMT.

In order to limit the number of the electronics channels
but keep high time resolution, we develop a read-out scheme using transmission lines as presented in this publication.
This read-out scheme is inspired by the approach developed in~\citep{Kim2012,Grabas_2013,Kim2016,Angelico_2017},
but has substantial differences in the way we plan to use it.
In references~\citep{Kim2012,Grabas_2013,Kim2016,Angelico_2017} the PMT is coupled with
high yield scintillation crystal (LYSO, LSO), thus
a large number of optical photons induce signals on all lines.
In such configuration, it is  not possible to distinguish signals from individual photons,
but rather reconstruct the mean position of the induced signals along and across lines.
In our approach, 1 to 2 photo-electrons are generated in average at
the photocathode~\citep{Ramos2015}. The main focus of our studies
is to optimize the whole system to have the best possible performance
for an individual photo-electron signal.

\section{Readout Realization}
\begin{figure}
    \centering\includegraphics[width=.36\textwidth]{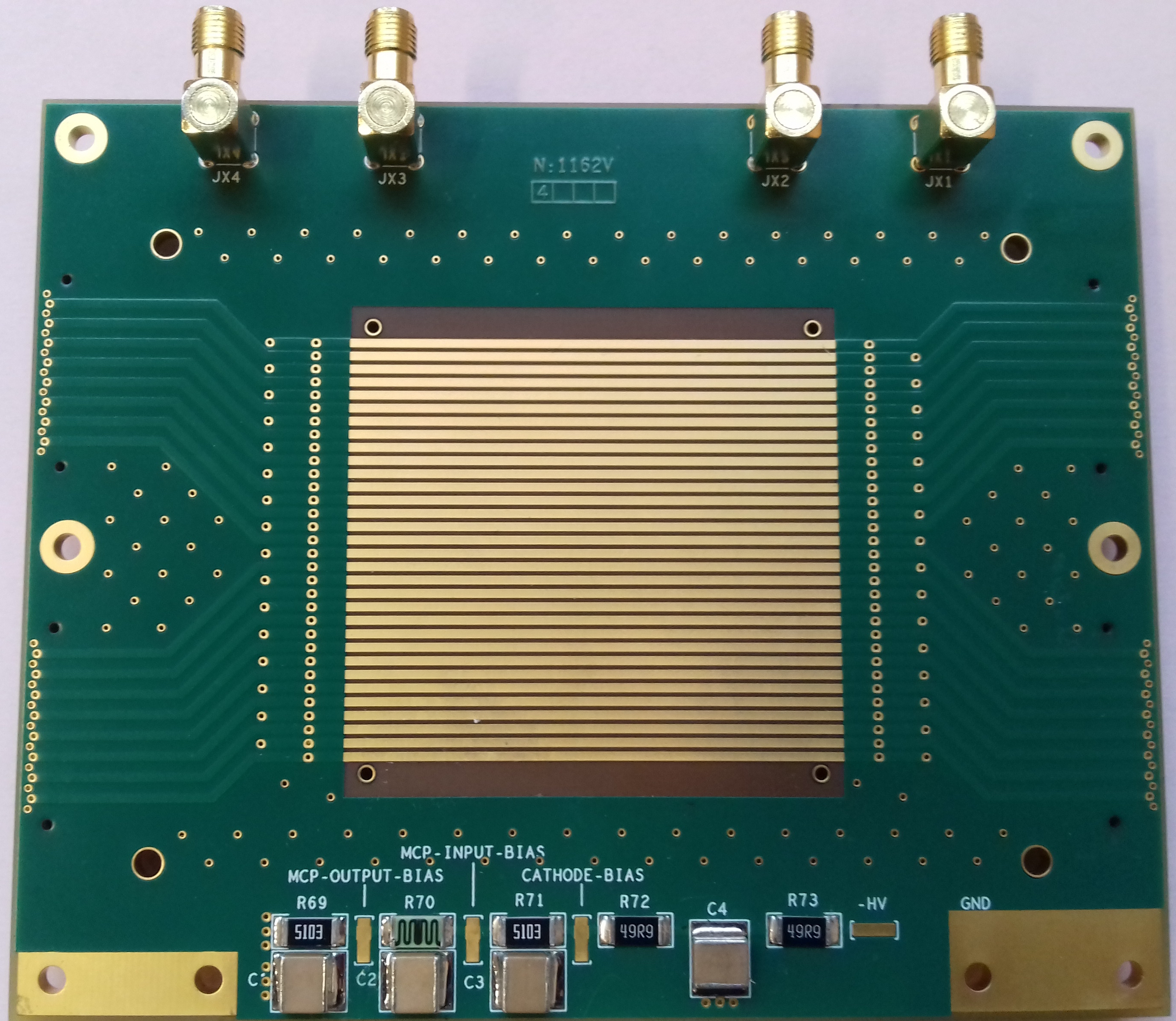}
    \caption{\label{fig:TL_PCB} Transmission lines PCB.}
\end{figure}

To study the feasibility and performance of transmission line read-out we use Photonis MCP-PMT Planacon XP85122 with
32 x 32 anode structure and 10~$\mu m$ pore diameter~\citep{XP85122_DataSheet}.
Each row of 32 pads (1.1 mm side, inter-pads distance is 0.5~mm)
is connected to a line at the printed circuit board (PCB) and signals induced at the line
are read-out from both ends, see Fig.~\ref{fig:TL_PCB}.

\begin{figure}
    \centering\includegraphics[width=.33\textwidth]{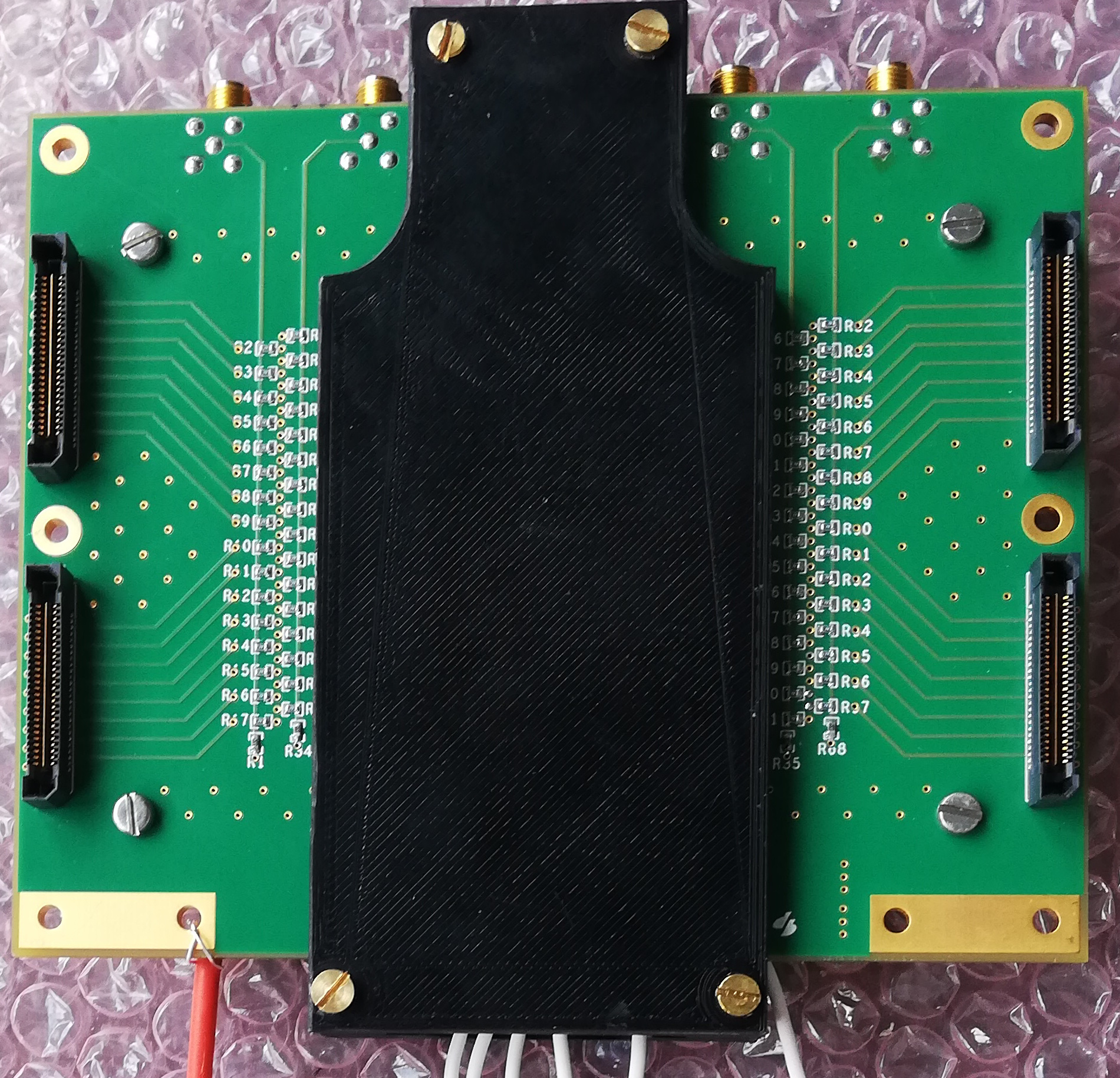}
    \caption{\label{fig:brace} Transmission line PCB clamping to the PMT with the plastic brace.}
\end{figure}

In order to be able to test different PCBs with the same PMT, 
we use the pressure-sensitive anisotropic conductive interface to connect transmission line  PCB to the
anode pads. Such reworkable solution doesn't require soldering and provides a low contact resistance, when
sufficient pressure is applied.
We tested two types of interfaces: the 3M\texttrademark\ adhesive tape \citep{ECATT} and
Shin-Etsu MT-type of Inter-Connector\textsuperscript{\textregistered} sheet \citep{Interconnect}. Both solutions
give reasonably good contacts under the pressure provided by the plastique or metallic brace, as shown in Fig.~\ref{fig:brace}.

The transmission lines board should ensure an electrical contact of matched impedance between
the PMT anodes and amplifiers.
We have chosen a PCB thickness of 3 mm to ensure sufficient
rigidity. It is also desirable that the transmission line protrudes
from the electronic board, in order to maximize the pressure
between contacts. We therefore removed the varnish in the contact area, and
used a copper thickness of 70 µm + gold plating, for a total
line thickness of 100 µm.  The transmission lines were carefully
matched to 50 Ohm, up to the connectors. This includes an
impedance calculation of the necessary vias, and careful
implementation of ground planes.  We placed the MCP-PMT high voltage
bias circuit (3~kV) on this board with a low pass filter at the input
to damp noises from the power supplies.  Finally, we made sure that
the line length were as homogeneous as possible, in order to avoid
excessive delays in the readout of the signals.

\begin{figure}
  \centering\includegraphics[width=.40\textwidth]{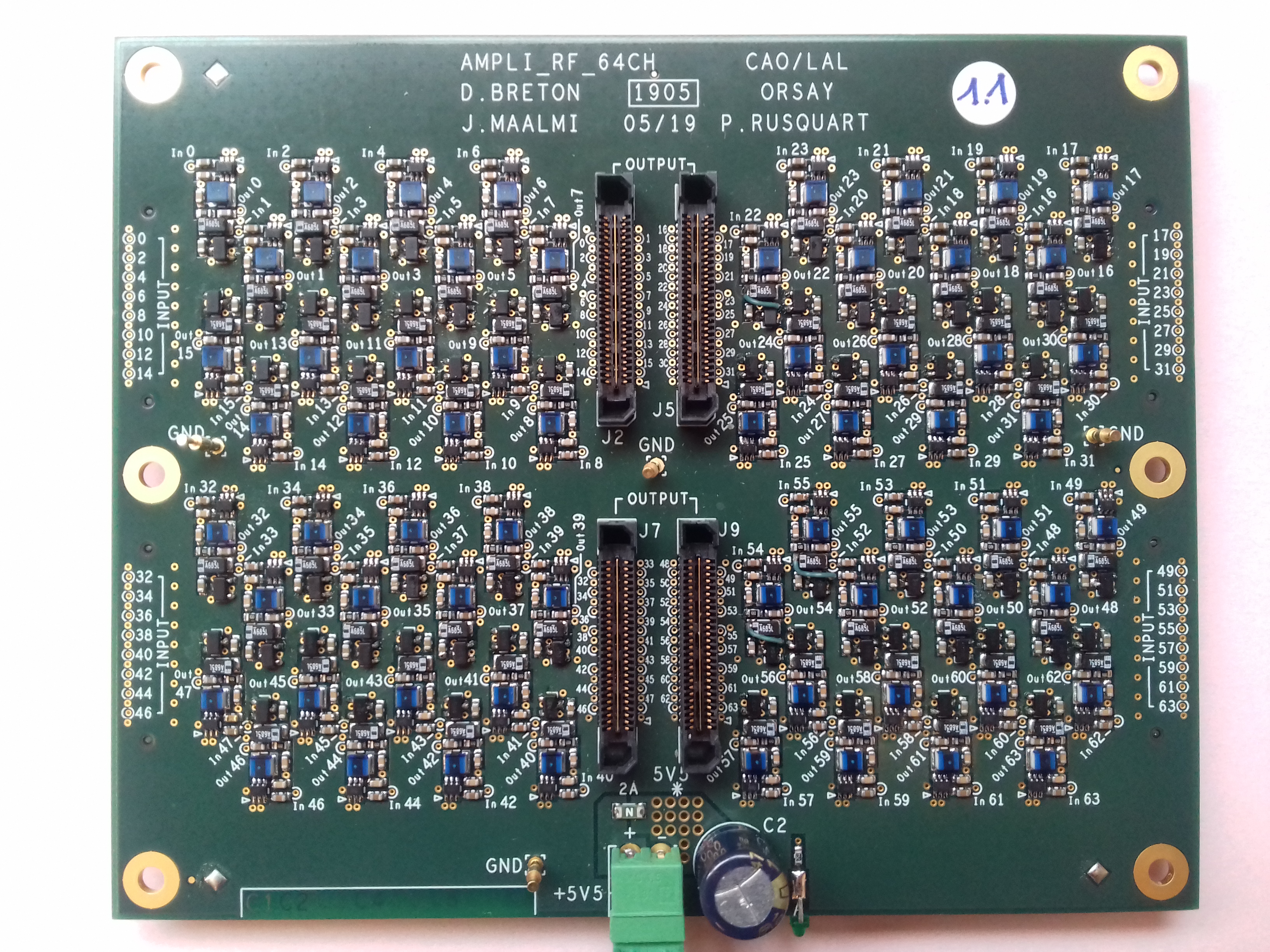}
\caption{\label{fig:AMPLI_PCB} Amplification board.}
\end{figure}

\begin{figure}
  \centering\includegraphics[width=.35\textwidth]{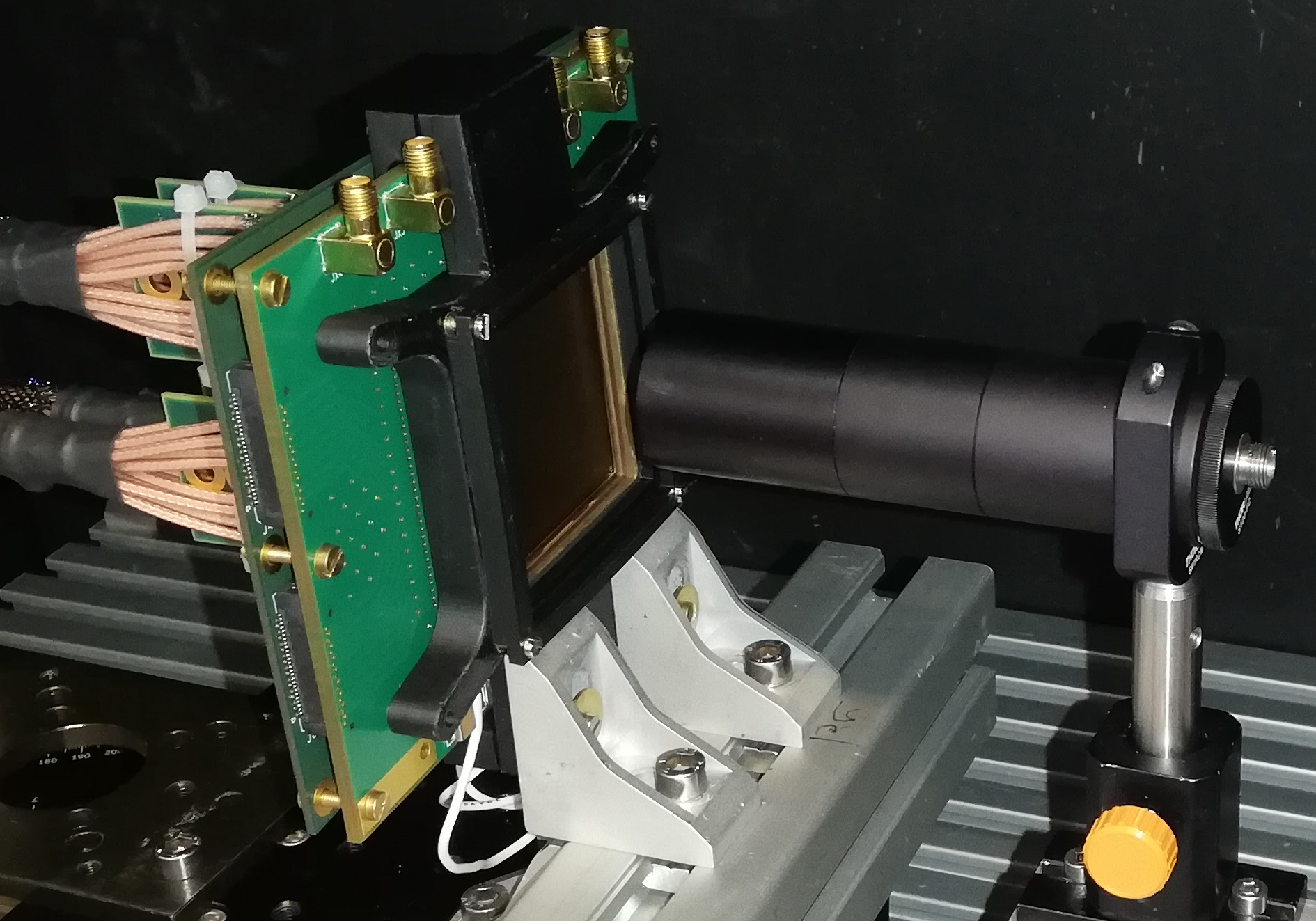}
  \caption{\label{fig:scan} PMT with the readout system, mounted in the black box.}
\end{figure}

The signals from both ends are amplified by 64  amplifiers, mounted on a PCB which
is directly plugged to the transmission line PCB (TL board) through
high density SAMTEC connectors QRM8/QRF8~\citep{Samtec},
Fig.~\ref{fig:AMPLI_PCB}.
The typical rise time of this PMT pulse is measured to be 450~ps (10\%--90\%), so 
we decided to limit the bandwidth of the amplification stages to 700 MHz to smooth slightly the signal.
For a 6.4~GSPS sampling rate of SAMPIC, this allows to capture three samples on the leading edge and avoid thus the
undersampling effect that could degrade the time calculation.
The single photo-electron signals at the output of
MCP-PMT are a few mV. A gain of 100 is therefore appropriate for the
1200 mV range of the SAMPICs input.
The integration of the 64 amplifier channels on an area close to that of the PMT
imposes severe integration constraints and also requires
to control thermal dissipation and crosstalk phenomena between
channels to ensure the stability and reliability of the boards.
Two amplification stages were necessary
for the required performance and to fully meet the specifications. Each
card, providing an amplification stage, dissipates 6W for 64 amplifier channels, a very low value in
this frequency range.
In this study, the first amplification stage is mounted directly on the PMT board,
and  the output signals from the amplifiers are transmitted through the 50 Ohm cables
to the second amplification stage outside of the black box.
The PMT, assembled with the TL board and 1st stage amplification board, is shown in Fig.~\ref{fig:scan}.

\begin{figure}[t]
\centering\includegraphics[width=.45\textwidth]{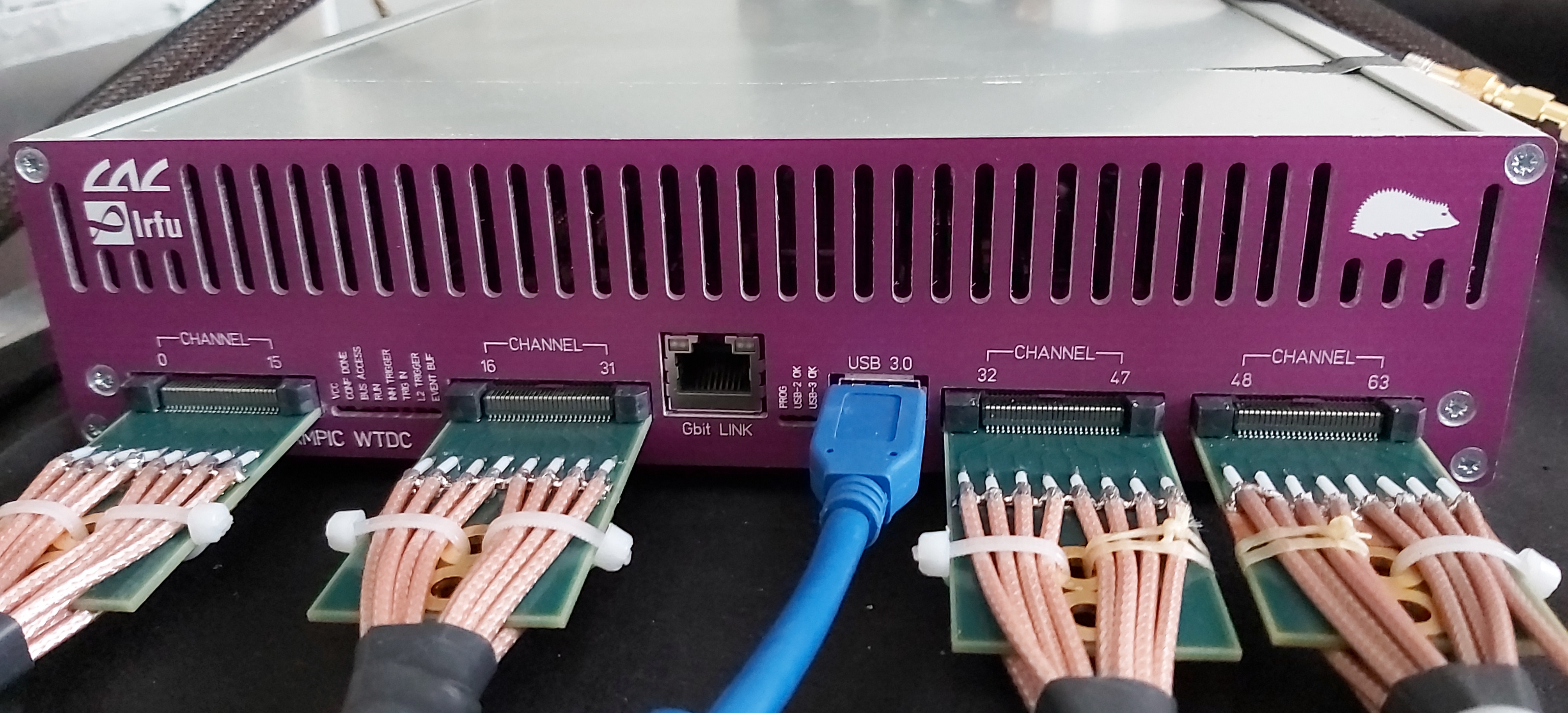}
\caption{\label{fig:SAMPIC} 64-channel SAMPIC module.}
\end{figure}

Finally, all signals are digitized by the 64-channel
SAMPIC module~\citep{Delagnes2014,BRETON2014,Delagnes:2015oda,Breton2020},
which uses four 16-channel SAMPIC\_V3C
chips, based on the patented concept of waveform and time-to-digital converter, Fig.~\ref{fig:SAMPIC}.
Each channel of the chip includes a delay-locked loop based TDC to provide a rough time
associated with an ultra-fast analog memory. The signal sampling frequency can be varied from
1.6 to 8.5 GS/s (6.4 GS/s frequency is adopted in this work).
The acquired waveforms are used for the precise timing measurement on-line,
but also can be stored for the off-line processing.
In this study, we are using a constant fraction discriminator algorithm (CFD) to estimate the time of a signal
with the threshold of 0.5 $\times$~amplitude.
This module allows to reach a timing accuracy of about 5~ps (SD) between channels from one chip,
and slightly larger values between different chips~\citep{Breton2016}.
Every channel integrates a discriminator that can trigger independently.
A first trigger level is implemented on-chip,
while a chip-related logic is realized at the module level using 4 inputs from 4 chips.

\section{Time and Spatial Resolution}
We test the designed system by scanning the surface of the PMT with the optical photons emitted by
a pulsed laser as described below.
The results of the scan allow us to measure a PMT transit time spread (TTS) at different positions
and investigate the spatial resolution of the individual photon reconstruction. 
Indeed, a difference in signal time from the left and right ends of the line allows to reconstruct the coordinate along the line (x-coordinate).
The sum of left and right signal times measures the detection time of a photon.
The coordinate across lines (y-coordinate) is reconstructed as a weighted mean of charge, integrated on each line.

\subsection{Measurement}
For the measurement  we use the pulsed laser Pilas by A.L.S.~\citep{Pilas}  as a light source.
The laser beam has a Gaussian-like time profile  with duration of about 20~ps (FWHM\footnote{FWHM: Full Width at Half Maximum. FWTM: Full Width at Tenth of Maximum}) and a jitter of 1.4 ps~\citep{Pilas}. 
The light beam from the laser fiber is collimated by a pin-hole  of 100~\um\ diameter.
The system of the laser fiber and pin hole is mounted at two-axis X-Z motion system, assembled from two \mbox{X-LRT0100AL-C} linear stages 
from Zaber Technologies Inc. This system allows to move and position the light source 
with a precision better than 25~\um~\citep{Zaber}.

The distance between laser output and pin-hole is about 100~mm and between pin-hole and PMT window is about 20~mm,
resulting in a light spot of diameter 120~\um.
We choose distances and the light intensity in such way that the PMT is working in a single-photon regime with 
a detection efficiency of 2\% corresponding to a ratio of two-photons~/~single-photon events  of 1\%.
This number is sufficiently small, that in the following studies we ignore the presence of events with two photons.
We realized a detector scan with step of 2~mm along the lines and 0.2~mm across the lines.
The data taking time of 1.5~s in each position allows to accumulate  of about 3000~events  per position.
For each signal we register amplitude and CFD  signal time  for all channels in coincidence with the laser trigger.
We use the so-called ``L2 coincidence'' option of the SAMPIC module, 
with a 30~ns coincidence time window.
The threshold value for accepting signals is 50~mV (a typical signal amplitude is about 300 -- 400~mV).

\subsection{Time Resolution}
\begin{figure}
  \begin{minipage}[t]{.23\textwidth} 
    \centering\includegraphics[width=\textwidth]{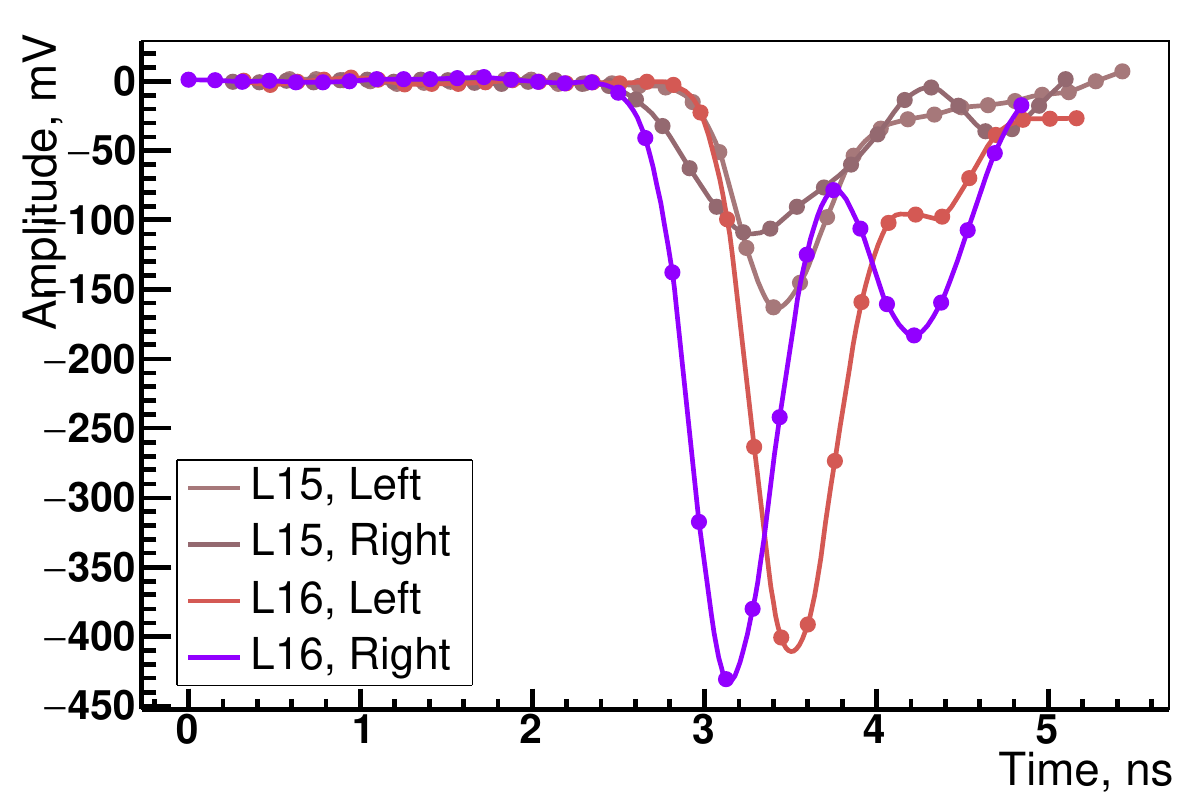}
  \end{minipage}
  \hfill
  \begin{minipage}[t]{.23\textwidth} 
    \centering\includegraphics[width=\textwidth]{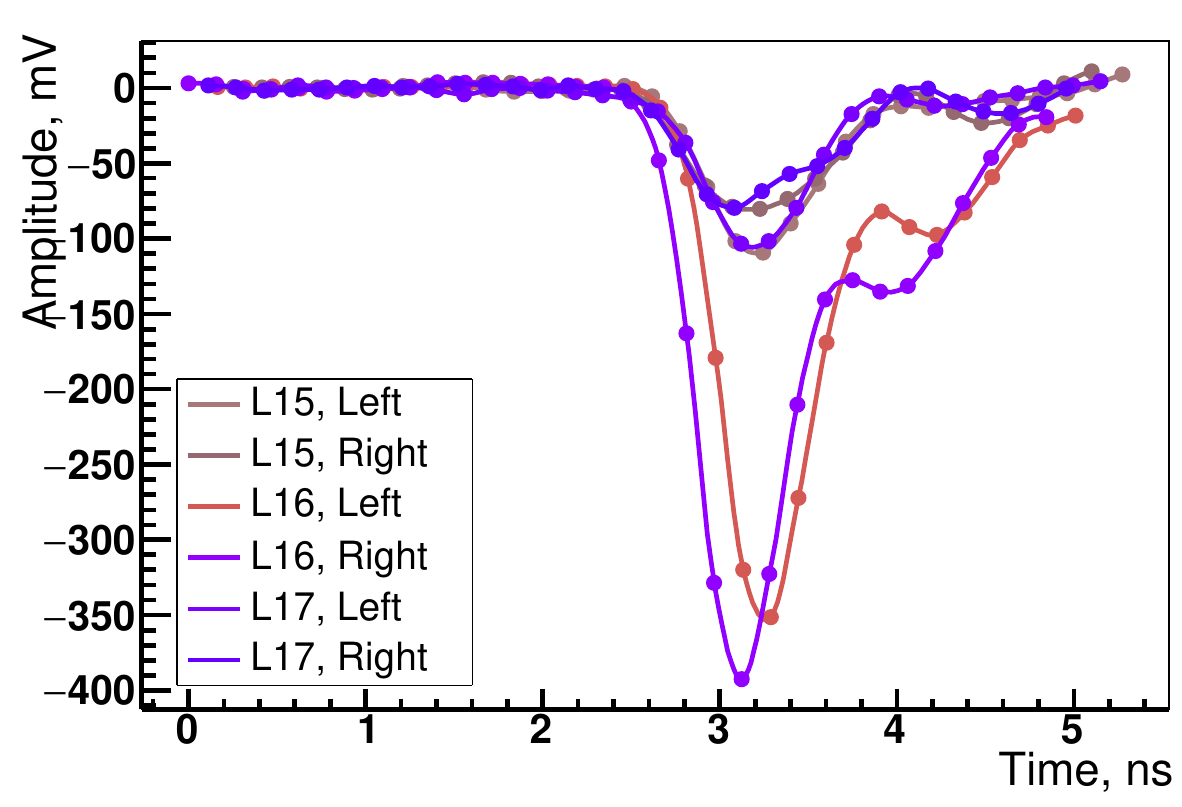}
  \end{minipage}
  \hfill
  \begin{minipage}[t]{.23\textwidth} 
    \centering\includegraphics[width=\textwidth]{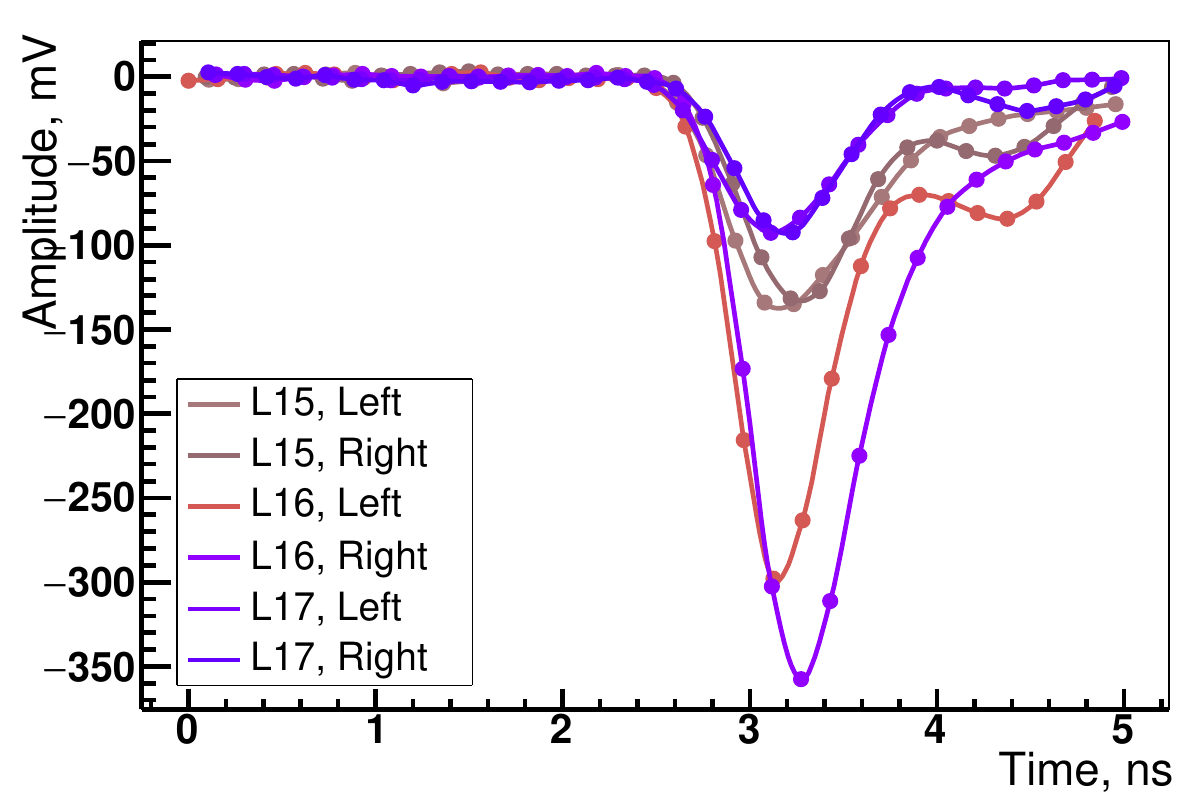}
  \end{minipage}
  \hfill
  \begin{minipage}[t]{.23\textwidth} 
    \centering\includegraphics[width=\textwidth]{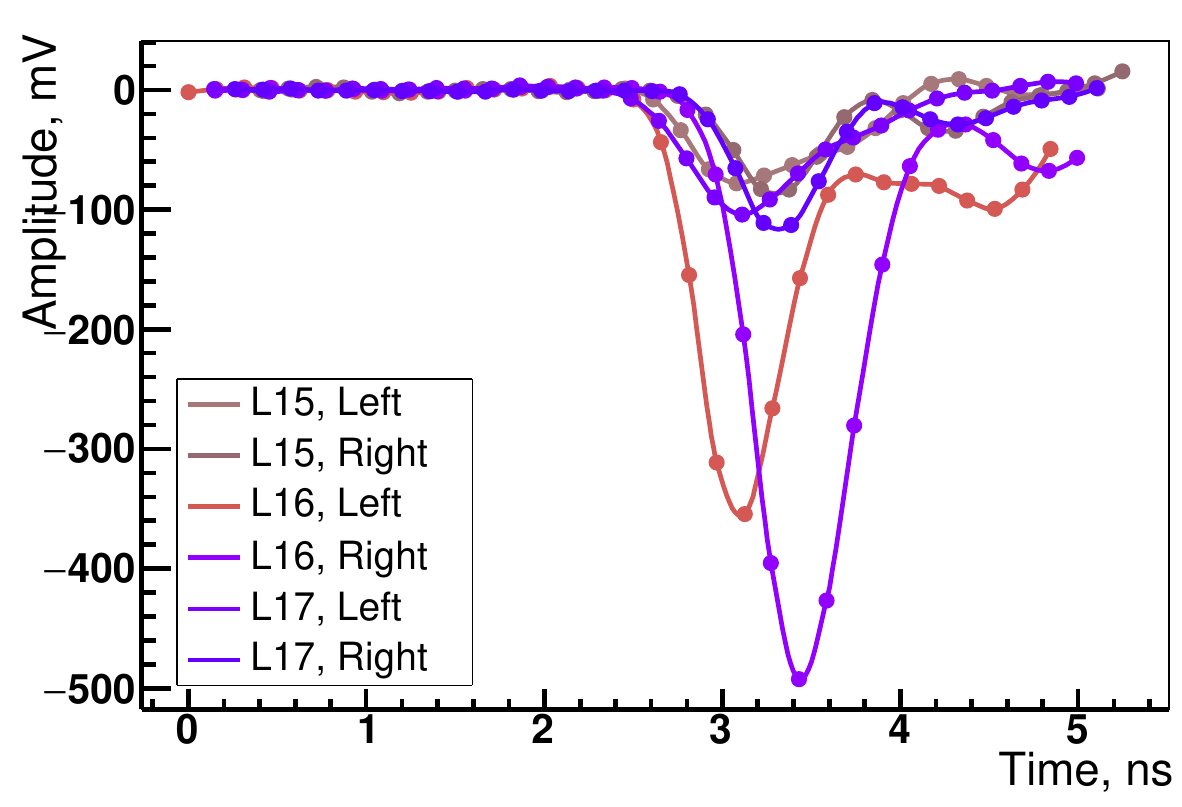}
  \end{minipage}
\caption{\label{fig:signals} Typical signals registered for positions (from left to right): (x, \mm; y, \mm) = (29.0,65.2) (41.0,65.2) (53.0,65.2) (65.0,65.2)}
\end{figure}

The typical signals are shown in Fig.~\ref{fig:signals} for different position along the same line.
As one can see, typically 2-3 lines are responding to the electron avalanche produced by the photon.
The reason for this is that during  the propagation between the MCP and the anode plane,
an electron avalanche produces a signal on the surface of a few millimeters in size.
The largest amount of charge is induced on the closest line
and, typically, 3 times lower charge is induced on the neighboring lines. 
The irregularity in the contact between transmission lines PCB and anode
may lead to an increased contact resistance between pads and the line and so redistributes more charge to neighboring lines.

\begin{figure}
  \centering\includegraphics[width=.34\textwidth]{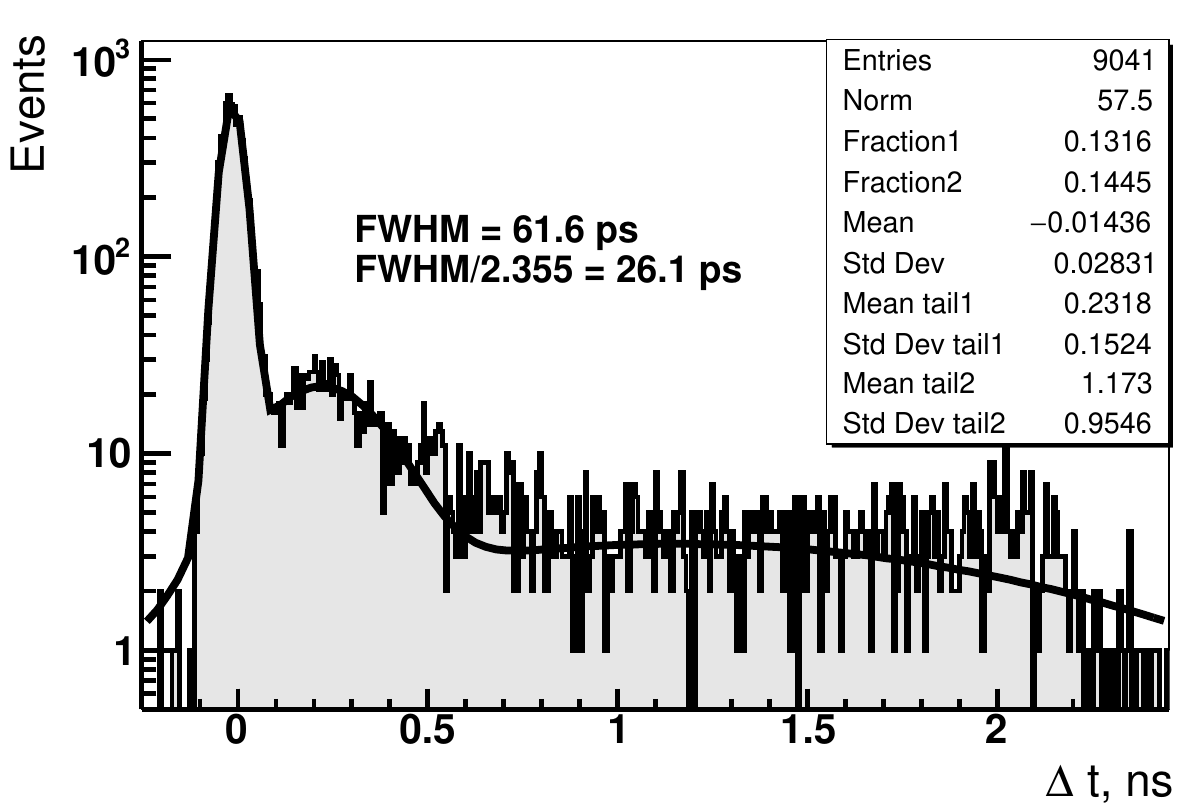}
    \caption{\label{fig:position_TTS} Difference in time between PMT and laser trigger signal for position (x, \mm; y, \mm) =
    (53,66).
    Statistical box shows the fit results with function~(\ref{eq:reso_func}).
    The FWHM is calculated by interpolation between neighboring bin values. }
\end{figure}

 To determine the time of arrival of a photon (PMT time),
 we choose the line with the highest amplitude and sum up  signal time measured at the left
 and right ends of the line. The difference between the PMT time and laser trigger time is
 shown in Fig.~\ref{fig:position_TTS}. We decided to fit this spectrum with a triple Gaussian formula:
\begin{multline}
  f(t) = \frac{n}{\sqrt{2\pi}} 
  \Bigl ( \frac{1-f_1-f_2}{\sigma_1}\ e^{-\frac{1}{2}(\frac{t-t_1}{\sigma_1})^2}   \\
  +  \frac{f_1}{\sigma_2}\ e^{-\frac{1}{2}(\frac{t-t_1-t_2}{\sigma_2})^2}  
   +  \frac{f_2}{\sigma_3}\ e^{-\frac{1}{2}(\frac{t-t_1-t_3}{\sigma_3})^2} 
   \Bigr ) \  ,
\label{eq:reso_func}
\end{multline}

where $n$ is a normalization coefficient, 
$f_1,\ f_2$ are fractions of events in second and third Gaussian terms, 
$t_1$ is the mean of the first term, 
$t_2,\ t_3$ are the additional delays for the second and third terms,
$\sigma_1$, $\sigma_2,\ \sigma_3$ are the corresponding standard deviations.
We fit the spectrum in the range of [-0.5 ns, 2.5 ns].

\begin{figure}
  \centering\includegraphics[width=.34\textwidth]{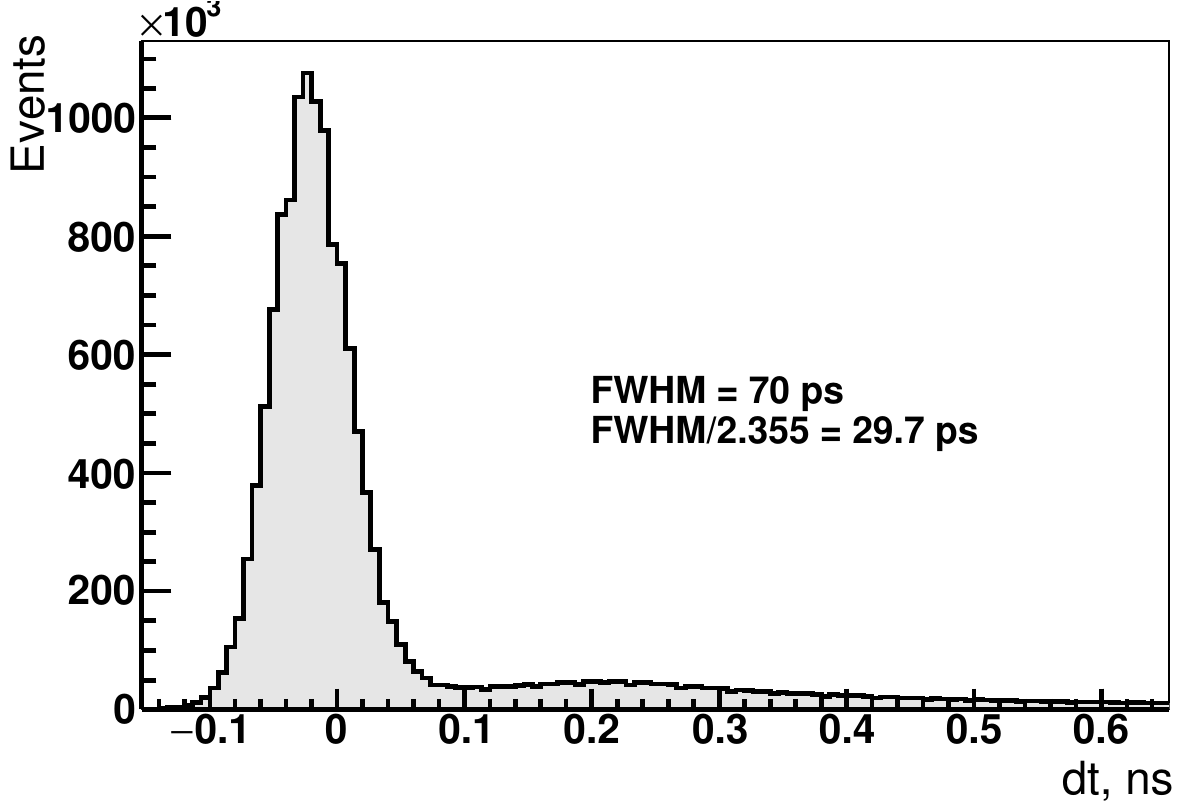}
    \caption{\label{fig:TTS_ref} Difference in time between PMT and laser trigger signal averaged over whole PMT surface.
      The FWHM is calculated by interpolation between neighboring bin values. }
\end{figure}

\begin{figure}
    \centering\includegraphics[width=.44\textwidth]{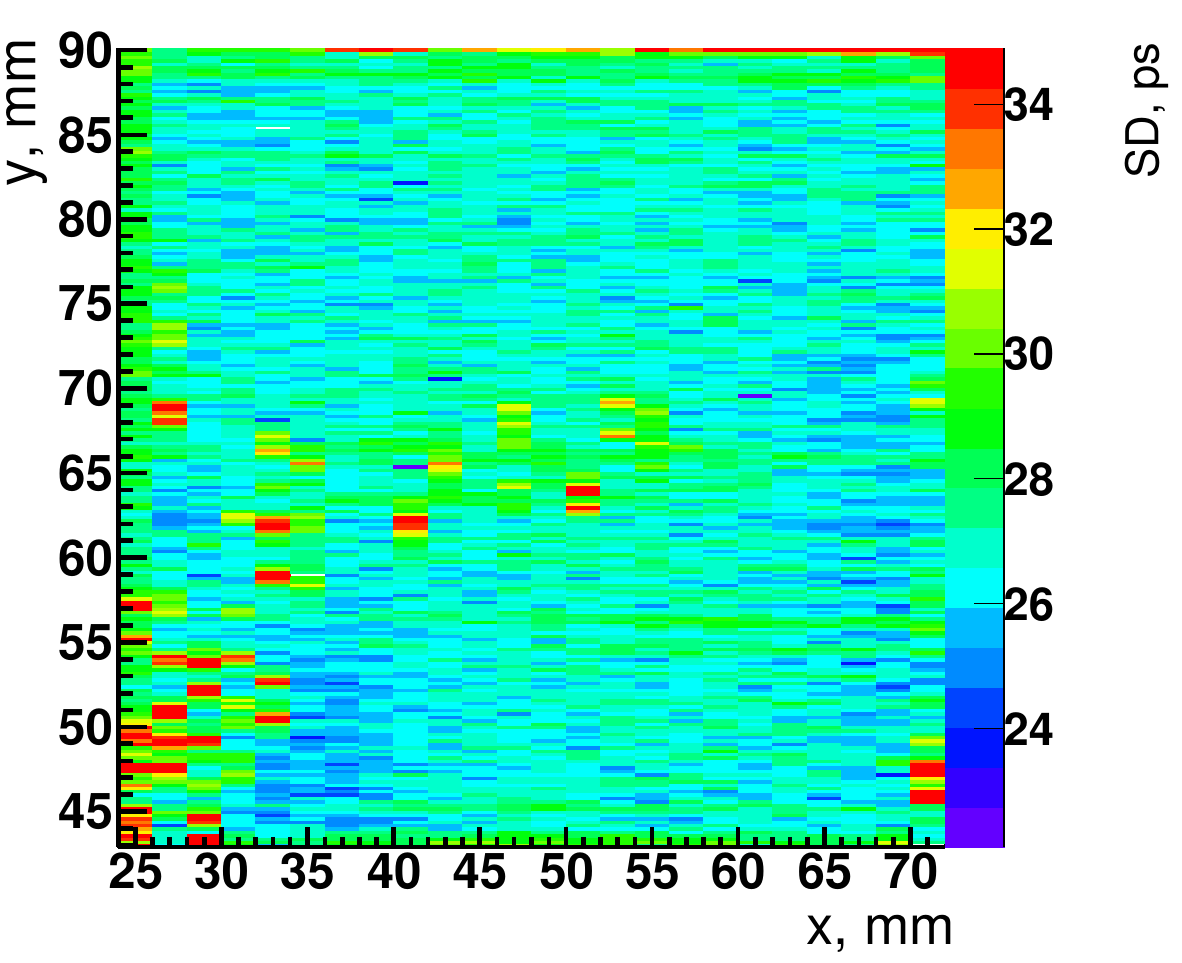}
  \caption{\label{fig:2D_width} Width (SD) of the time distribution (parameter $\sigma_1$ in the function~(\ref{eq:reso_func})).}
  \end{figure}

\begin{figure}[t]
    \centering\includegraphics[width=.44\textwidth]{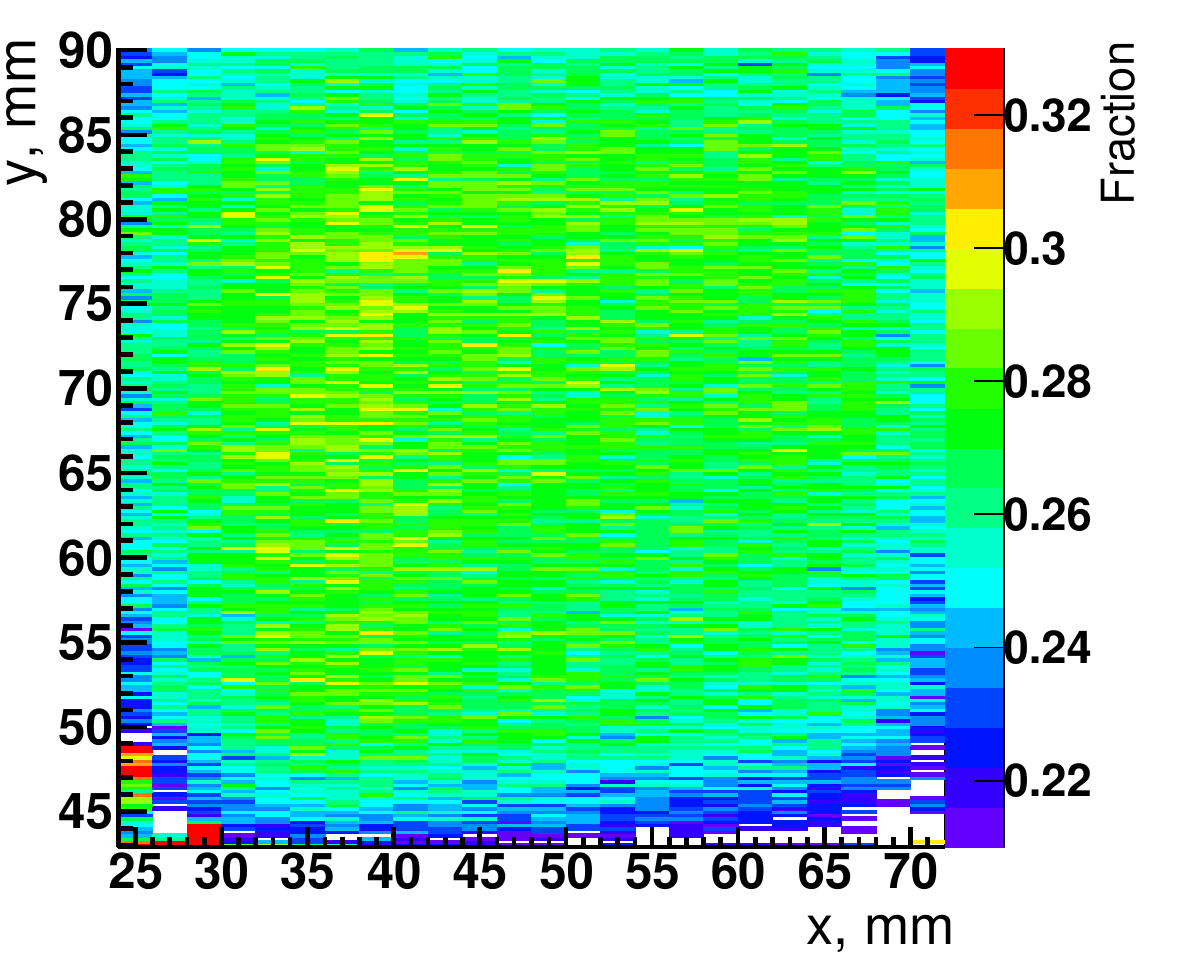}
  \caption{\label{fig:2D_fraction} Fraction of events delayed more than 100~ps relatively to the position of the main peak.}
\end{figure}

As one can see about 75\% of events have gaussian-like distribution with the width of about 67~ps (FWHM) and the other
25\% corresponds to the back-scattered electrons~\citep{Korpar2008Sep,Lipka2018} and form a long tail. 
Such fit is done in each position and Fig.~\ref{fig:TTS_ref} shows the value averaged over whole PMT surface.
The obtained value 70~ps (FWHM) is larger than the simple average in all positions, 65~ps, due to the residual imperfections  in calibration.
Fig.~\ref{fig:2D_width} and \ref{fig:2D_fraction} show the width of the distribution as well as
the fraction of electrons delayed for more than 100~ps as a function of x and y-coordinates.

\subsection{Spatial Resolution}
\begin{figure}[t]
  \centering\includegraphics[width=.44\textwidth]{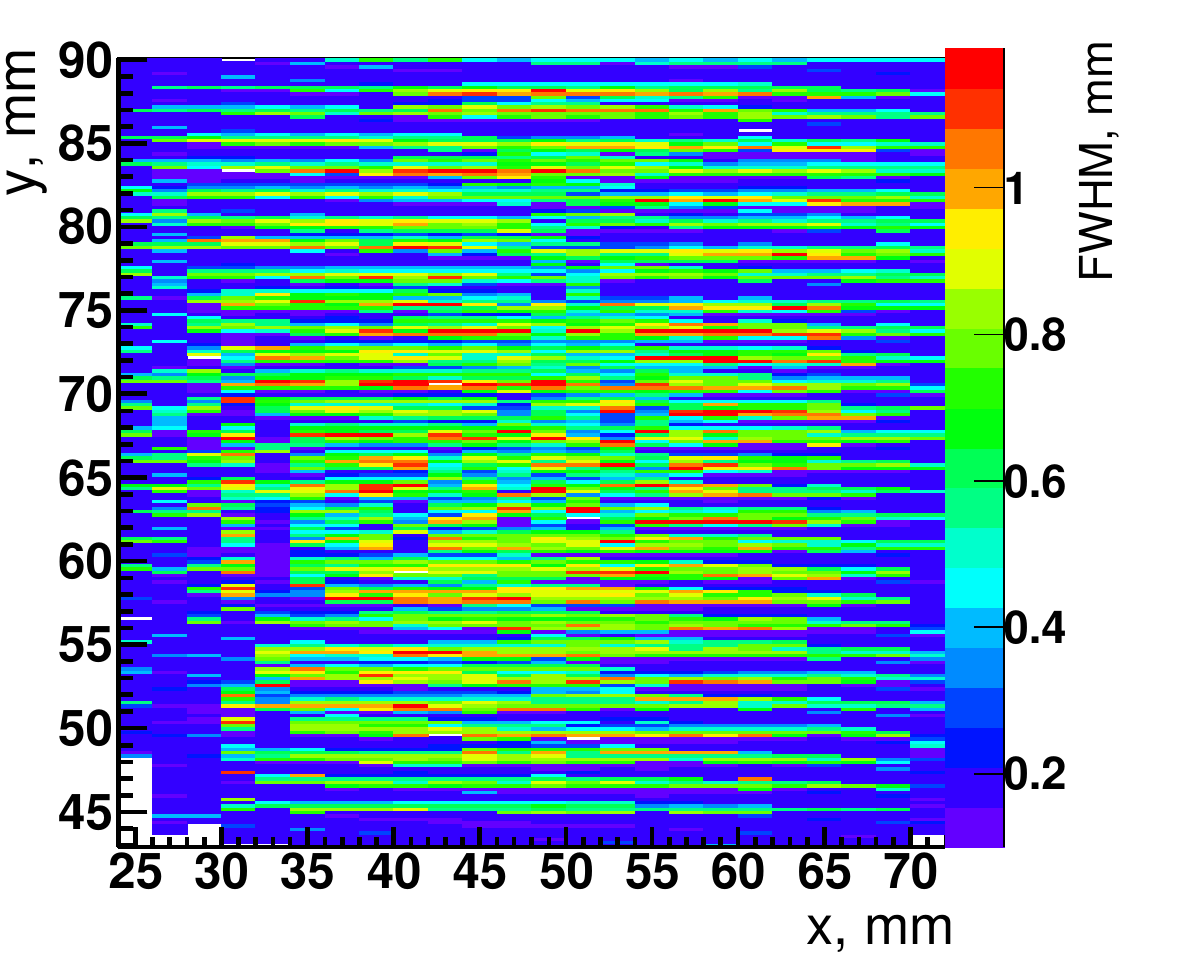}
  \caption{\label{fig:dy_2d} Resolution of the reconstructed y-coordinate (across lines, FWHM) as a function of
     linear stage's x- and y-coordinates.}
  \end{figure}

\begin{figure}
    \begin{minipage}[t]{.23\textwidth}
      \centering\includegraphics[width=\textwidth]{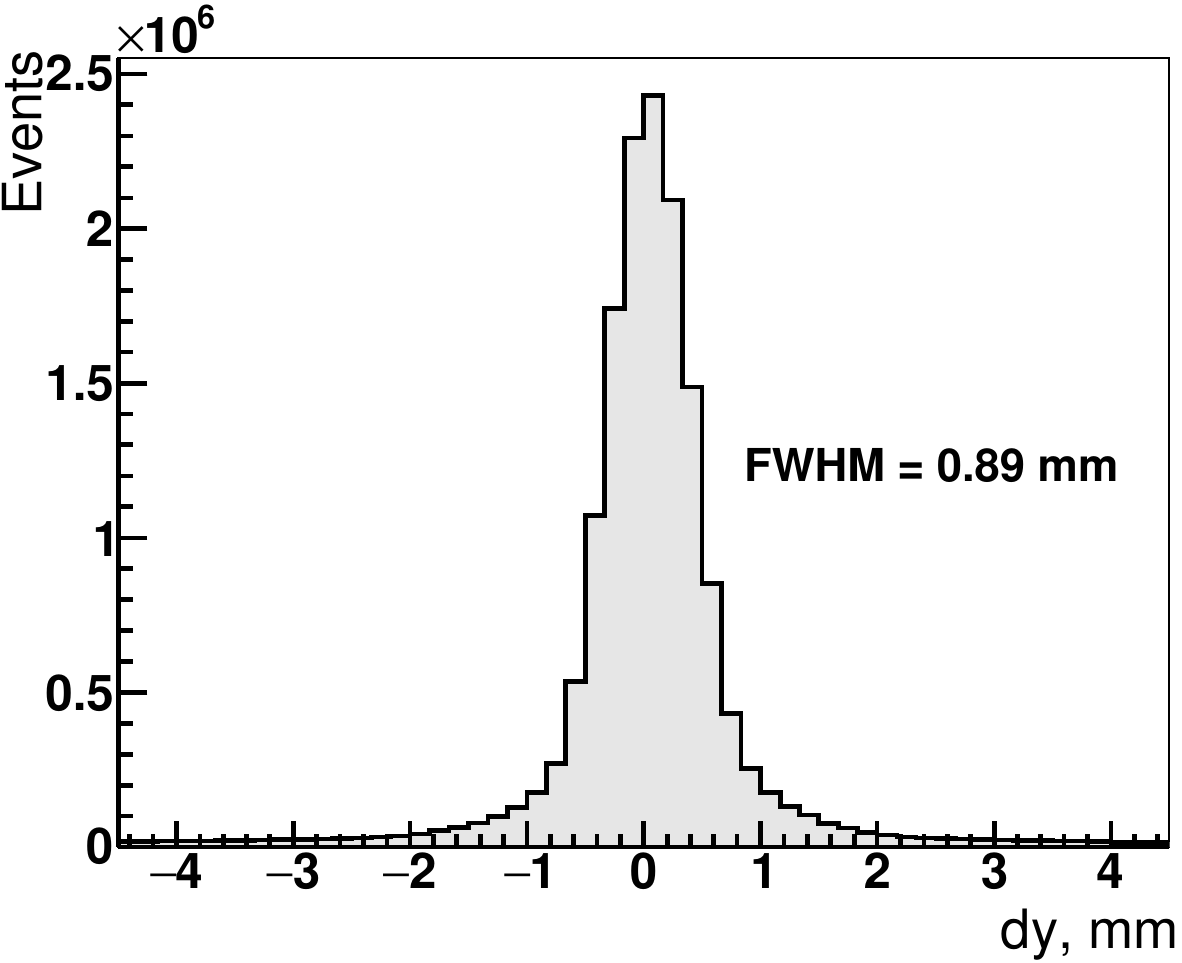}
  \caption{\label{fig:dy_1d} Difference between reconstructed and  linear stage's  y-coordinates for all positions. }
    \end{minipage}
    \hfill
  \begin{minipage}[t]{.23\textwidth} 
    \centering\includegraphics[width=\textwidth]{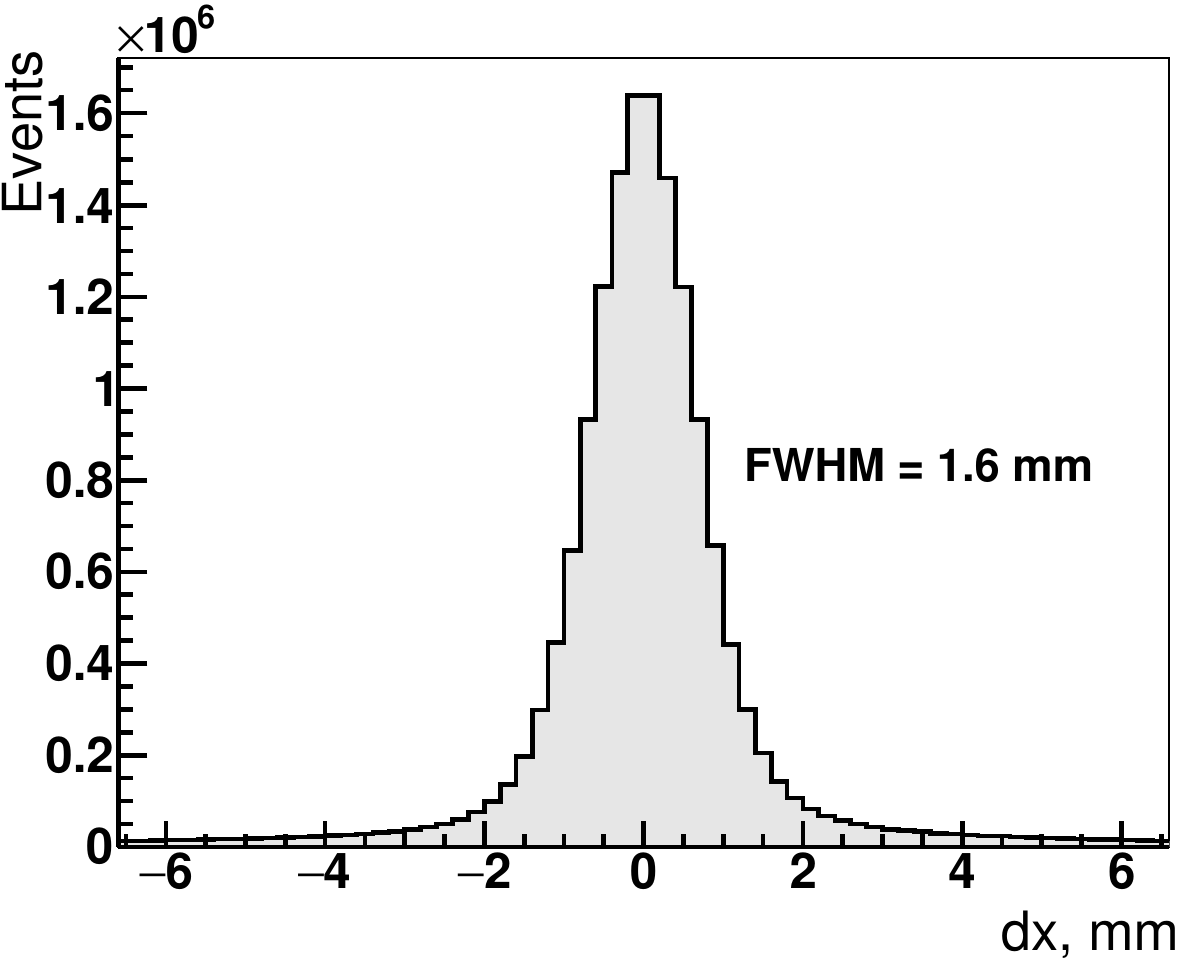}
    \caption{\label{fig:dx_1d} Difference between reconstructed and  linear stage's x-coordinates for all positions. }
    \end{minipage}
\end{figure}

To reconstruct the coordinate across lines ($y_R$), we choose the line with the largest amplitude and calculate
the weighted average of coordinates for this line and the two closest neighbors:
\begin{equation}
  y_R = \frac{\sum_{i=1}^3 y_i C_i}{\sum_{i=1}^3 C_i}\ ,
  \label{eq:y_reco}
\end{equation}
where $y_i$ is a y-coordinate of the center of line $i$, $C_i$ is a charge at the same line, measured as a sum of signals over all 64 sampling points.
Fig.~\ref{fig:dy_2d} shows the resolution of the reconstructed y-coordinate as a function of the linear stage's  x- and y-coordinates and
Fig.~\ref{fig:dy_1d}  shows the difference between reconstructed and linear stage's y-coordinates for all registered positions.
In average, the resolution of about 0.9~mm is achieved with a variation between 0.3 and 1.2~mm depending on the position.

\begin{figure}[t]
    \centering\includegraphics[width=.44\textwidth]{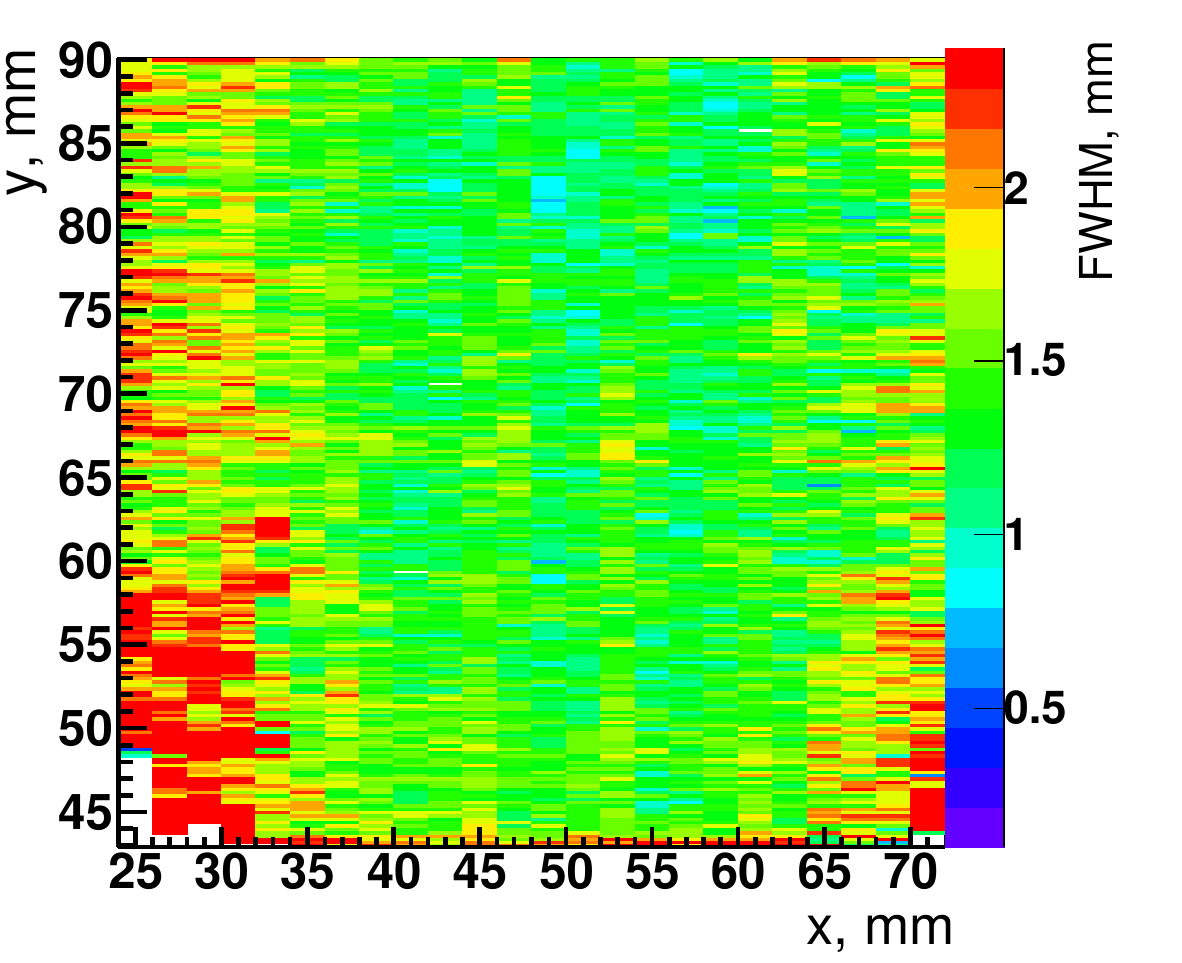}
    \caption{\label{fig:dx_2d} Resolution of the reconstructed x-coordinate (along lines, FWHM) as a function of
       linear stage's x- and y-coordinates.}
  \end{figure}

\begin{figure}
  \begin{minipage}[t]{.23\textwidth} 
  \centering\includegraphics[width=\textwidth]{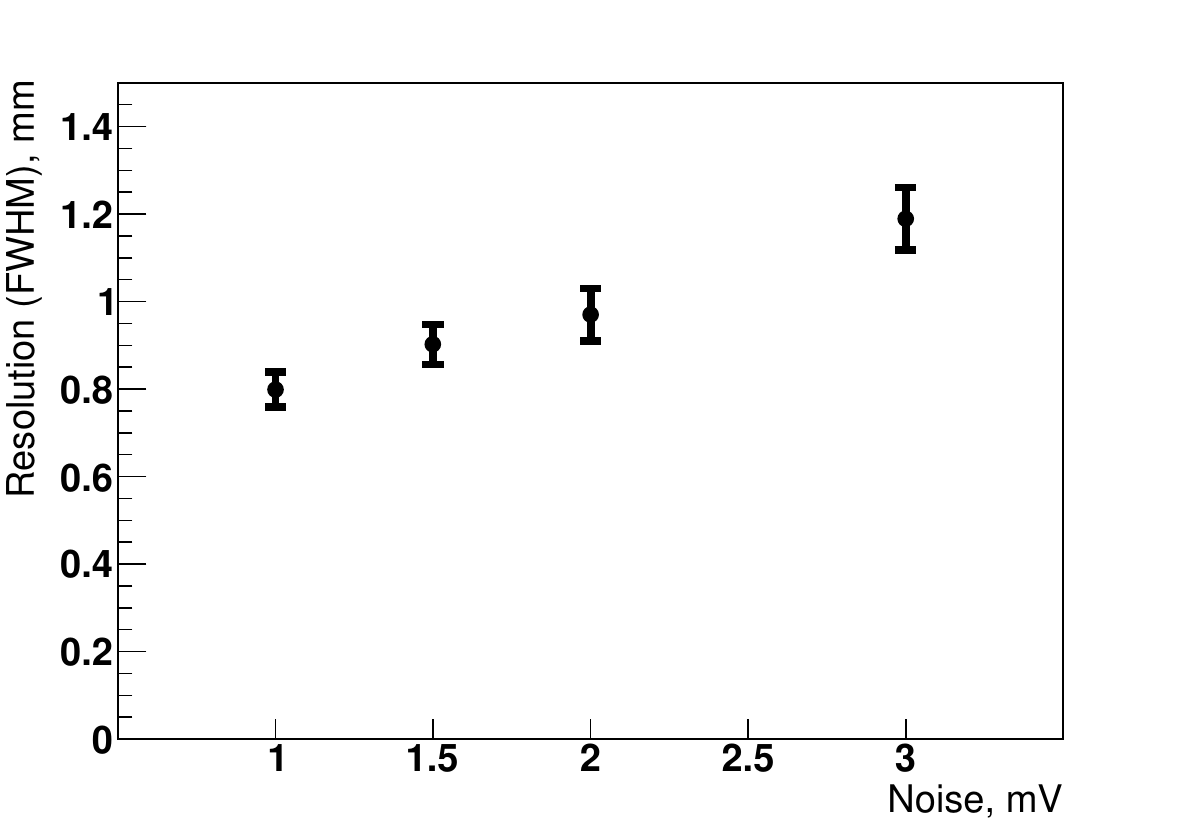}
  \caption{\label{fig:sim_res} Resolution of the x-coordinate reconstruction estimated by the simulation as a function
  of the electronic noise level.}
\end{minipage}
\hfill
  \begin{minipage}[t]{.23\textwidth} 
    \centering\includegraphics[width=\textwidth]{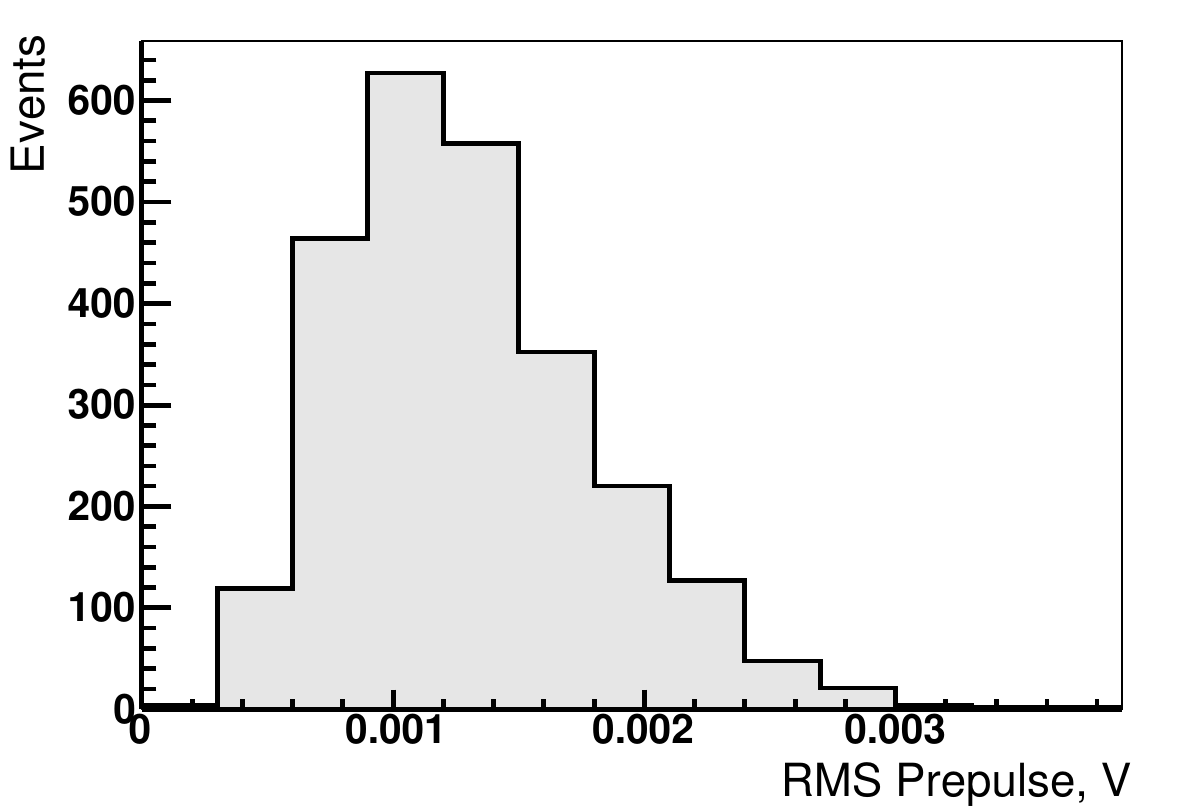}
    \caption{\label{fig:noise} Typical noise distribution for a single position, measured as the RMS of the signal baseline samples. }
\end{minipage}
\end{figure}

The coordinate along lines ($x_R$) is reconstructed as $x_R = (t_R-t_L) s$, where $t_R$ and $t_L$ is a time measured at the right and left line extreme
respectively and $s$ is a signal propagation speed, measured to be about 0.36 x speed--of-light. 
Fig.~\ref{fig:dx_2d} and \ref{fig:dx_1d} show the resolution distributions for the x-coordinate. We obtain a precision of about 1.6~mm
(FWHM) in average with the variation between 0.8 and 2.5~mm.
The resolution in x depends on the precision in the signal time evaluation  at  both ends of the line.
In order to estimate the expected precision, we developed a simulation, which uses
the measured shape of the signal, simulates the digitization of signals with the step 156.25~ps,
applies the randomized white noise to the digitized shape.
Fig.~\ref{fig:sim_res} shows the estimated time resolution calculated for the different levels of the electronics noise.
The noise in the experiment is measured using signal baseline and includes all electronics contribution,
e.g. noise during amplification, propagation in cables, digitization by the SAMPIC module.
As shown, in Fig.~\ref{fig:noise} the typical value is about 1~mV and hence the expected spatial resolution
along the line should be below 1~mm according to the simulation.
The lowest value observed in the measurement corresponds roughly to the value in the simulation.
The average resolution value of 1.6 mm is significantly larger due to effects not accounted for
in the simulation, e.g. signal shape deformation during the propagation over the transmission lines,
fluctuation in the signal shape and amplitude related to the uniformity of the PCB,  non-uniformity of the electrical contacts between PCB and pads,
border effects, imperfection in calibration functions etc.

\subsection{Discussion}

Results presented in the previous sections demonstrate that one can reconstruct a single-dielectron coordinate
along and across the readout line with a precision of 1.6~mm and 0.9~mm simultaneously with the time resolution of about 70~ps.
This time resolution is limited mainly by the PMT TTS with a small contribution  due to imperfections
in the calibration.
For the simultaneous  detection of two and more photons the situation will be quite different.

Registered photo-electrons separated by at least two lines,
will be reconstructed independently and with the same precision.
On the contrary, if photo-electrons are happened to be on the same line, it will lead to the overlap
between the induced signals and to a deformation of the signal
shape.
In the BOLD-PET projects, we expect to detect in average 1 -- 2 optical photons per event~\citep{Ramos2015}.
The probability that those two photons will cause the signal overlap is quite low,
and hence such events could be rejected.

Such read-out can be and is planned to be used in the situation with a larger number of detected photo-electrons.
For example, we plan to use similar read-out scheme in the ClearMind
project~\citep{Yvon2020Jul}, where the expected number of photo-electrons per event is about twenty.
Depending on the depth-of-interaction of the gamma conversion,
the detected events may contain more or less overlapping signals.
In such situation, more complicated reconstruction algorithms and, in particular,
algorithms based on the machine learning approach,
allow to mitigate the overlap and
reconstruct the gamma-conversion coordinates with a millimeteric precision without degrading the time resolution.

\section{Conclusion}
In this study we demonstrate the possibility to read-out the MCP-PMT with transmission lines and obtain
simultaneously an excellent time resolution on whole surface of the PMT and good spatial resolution.
The obtained values fully satisfy requirements to the optical  read-out in the BOLD-PET project and will be 
used there to detect the Cherenkov light, generated by the conversion of the 511~keV gamma in the liquid TMBi radiator.

The use of amplifier boards and SAMPIC module, developed
in our labs, allow us to realize the cost-effective, multi-channel digitization of signals with excellent precision.
For a single photo-electron, we  obtain the average time resolution of 70~ps (FWHM), and
coordinates resolution of 1.6~mm and 0.9~mm (FWHM) along and across the line respectively.

\section*{Acknowledgments}
We thank professor Henry Frisch from the University of Chicago for the
fruitful discussion on the readout PCB  technology and
for providing us a transmission lines read-out board~\citep{Kim2012},
which was used for first tests in our development.
We thank Irakli Mandjavidze from IRFU, CEA for his studies and advises on optimizing signal propagation through transmission lines.

We acknowledge the financial support by the joint French-German grants  ANR-18-CE92-0012-01, DFG-SCHA 1447/3-1 and WE 1843/8-1.
This work is conducted in the scope of the IDEATE International Associated Laboratory (LIA).

\bibliography{References}

\end{document}